\documentclass[a4paper]{article}

\usepackage[margin=1in]{geometry} 

\usepackage{amsmath}
\usepackage{amsthm}
\usepackage{amssymb}
\usepackage{algorithm}
\usepackage{algorithmicx}
\usepackage{algpseudocode}
\usepackage{amssymb}
\usepackage{mathtools}
\usepackage{float}
\DeclarePairedDelimiter\abs{\lvert}{\rvert}%
\DeclarePairedDelimiter\bra{\langle}{\rvert}
\DeclarePairedDelimiter\ket{\lvert}{\rangle}

\usepackage{subcaption}

\usepackage[utf8]{inputenc}
\usepackage{hyperref}
\hypersetup{
	unicode,
	pdfauthor={Alice Barthe},
	pdftitle={Bloch Sphere Binary Trees: A method for the visualization of sets of multi-qubit systems pure states},
	pdfsubject={Bloch Sphere Binary Trees: A method for the visualization of sets of multi-qubit systems pure states},
	pdfkeywords={article, template, simple},
	pdfproducer={LaTeX},
	pdfcreator={pdflatex}
}

\usepackage[sort&compress,numbers,square]{natbib}
\theoremstyle{plain}

\theoremstyle{definition}

\usepackage{graphicx, color}
\graphicspath{{fig/}}

\usepackage{algorithm, algpseudocode} 
\usepackage{mathrsfs} 

\usepackage{lipsum}

\title{Bloch Sphere Binary Trees: A method for the visualization of sets of multi-qubit systems pure states}
\author{Alice Barthe$^{1,2}$, Michele Grossi$^1$, Jordi Tura$^2$, Vedran Dunjko$^2$}

\date{
	$^1$ Quantum Technology Initiative, CERN\\%
	$^2$ Applied Quantum Algorithms, Leiden University\\[2ex]%
}

\begin{document}
	\maketitle
	
	\begin{abstract}
        Understanding the evolution of a multi-qubit quantum system, or elucidating what portion of the Hilbert space is occupied by a quantum dataset 
        becomes increasingly hard with the number of qubits. In this context, the visualisation of sets of multi-qubit pure quantum states on a single image can be helpful. However, 
        the current approaches to visualization of this type only allow the representation of a set of single qubits (not allowing multi-qubit systems) or a just a single multi-qubit system (not suitable if we care about sets of states), sometimes with additional restrictions, on symmetry or entanglement for example. \cite{bib1,bib2,bib3}.
        
        
        In this work we present a mapping that can uniquely represent a set of arbitrary multi-qubit pure states on what we call \textit{a Binary Tree of Bloch Spheres}. The backbone of this technique is the combination of the Schmidt decomposition and the Bloch sphere representation. We illustrate how this can be used in the context of understanding the time evolution of quantum states, e.g. providing immediate insights into the periodicity of the system and even entanglement properties. We also provide a recursive algorithm which translates from the computational basis state representation to the binary tree of Bloch spheres representation. The algorithm was implemented together with a visualization library in Python released as open source.
		
	\end{abstract}

	\tableofcontents

 \pagebreak
 
\section{State of the Art}\label{sec1}

The most widely adopted representation of qubit based systems is the Bloch sphere which is a unique \footnote{NB "unique" or "bijective" in this work will always be up to the global phase.} mapping between a point on a unit sphere and a single qubit quantum state, allowing to represent several single qubit pure states on a single image. IBM’s q-sphere \cite{bib1} allows for the representation of a single multi-qubits pure state, by a set of colored points on a sphere corresponding to the computational basis decomposition. Koczor et al \cite{bib2} presented a representation for a specific family of many-body states (only permutationally symmetric states) as an intensity map on a sphere using spherical phase spaces. P. Migdał PhD thesis \cite{bib3} focuses on the representation of the same family of states and along with other methods, presents a method applicable to arbitrary multi-qubit states as a grid corresponding to coefficient in  basis where the color represent the phase and intensity the absolute value. All the fore mentioned techniques either allow several one qubit states representation either only one multi-qubit state on a single figure. In contrast the technique presented in this work allows for the representation of several arbitrary multi-qubit states on a single figure.

\begin{figure}[h]
\centering
\includegraphics[width=0.9\linewidth]{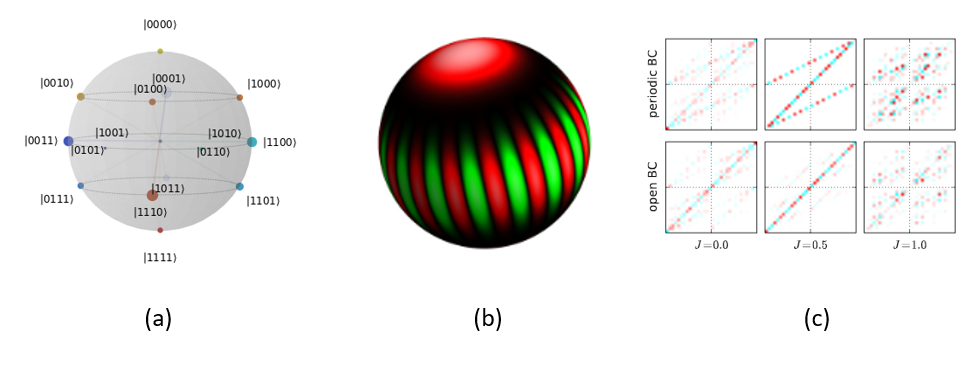}
\caption{ (a) 4 qubits entangled state qsphere \cite{bib1} (b) GHZ state spherical phase spaces \cite{bib2} (c) Majumdar-Ghosh model qubism representation \cite{bib3}}\label{fig1}
\end{figure}

\section{Method Description}\label{sec2}

The backbone of the technique presented in this work is the combination of the Schmidt decomposition and the Bloch sphere mapping. Given Hilbert spaces $H_A$ and $H_B$ corresponding respectively to $1$ and $N-1$ qubits, for any arbitrary pure state in the tensor product of two Hilbert spaces $ H = H_A \bigotimes H_B$ there exist a unique decomposition into two qubits in $H_B$ and two real angles $(\theta,\phi)$ such that: 
\[ \ket{\psi}_{AB} = \cos(\theta/2) \ket{0}_A \ket{\psi_0}_B + \sin(\theta/2) e^{i \phi} \ket{1}_A \ket{\psi_1}_B \]

This principle can be applied recursively until $H_B$ corresponds to only one qubit, in which case the recursion is stopped with the standard Bloch Sphere mapping. In other words, we apply nested Schmidt decomposition where each node can be represented as a Bloch sphere, therefore it can be decomposed into what we called a \textit{Binary Tree of Bloch Sphere (BTBS)}. We present a set of properties of this decomposition:

\begin{itemize}
  \item Applicable to sets of arbitrary multi-qubit system states (no required properties)
  \item The transformation is bijective up to the global phase.
  \item States that are close in the Hilbert space will be close in the BTBS.
  \item Product states can be identified visually
\end{itemize}

\begin{figure}[h]
\centering
\includegraphics[width=0.9\linewidth]{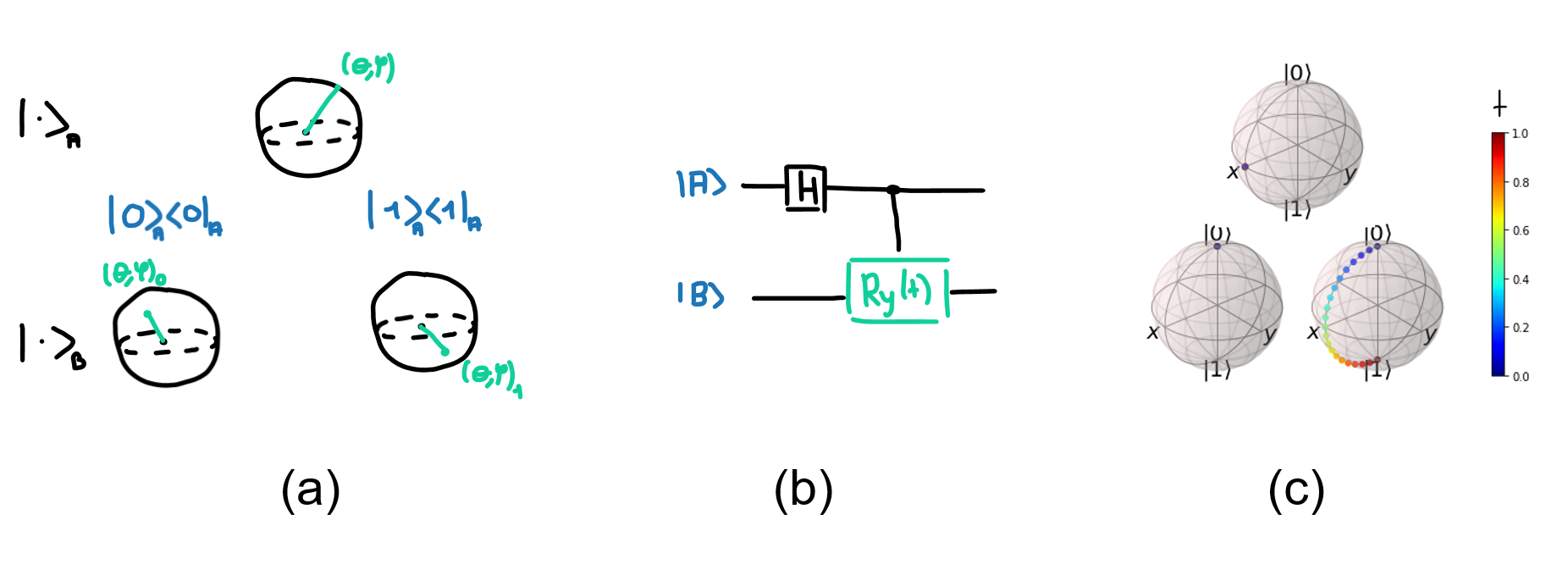}
\caption{ (a) schematic of 2-qubit BTBS (b) Parameterized Bell state preparation circuit (c) Parameterized bell state evolution on the BTBS  }\label{fig2}
\end{figure}

To illustrate the concept we present a simple circuit on two qubits as described on Fig.\ref{fig2}. It is a parameterised Bell state preparation, where the CNOT gate is replaced with a parameterized Y rotation gate. The first row represents the first qubit and is composed of a single sphere, as the Y rotation parameter changes, it remains in the $\ket{+}_A$ state. The second qubit is represented by the two spheres on the second row. The left sphere shows the second qubit in the subspace where the first qubit is projected onto $\ket{0}_A\bra{0}_A$, it remains unchanged on $\ket{1}_B$ because the controlled rotation only has effect on the subspace $\ket{1}_A\bra{1}_A$. The right sphere shows the second qubit in the subspace where the first qubit is projected onto $\ket{1}_A\bra{1}_A$, it starts from $\ket{0}_B$ (t=0, blue point) and then gradually rotates about the Y axis to end on $\ket{1}_B$ (t=1, red point). This also illustrates how product state can be identified. If two sub trees are the same (second row, blue points) then the state is a product state in the subspace corresponding to the binary tree coordinates. We can write the full Schmidt-Bloch decomposition for two qubits: 

\begin{equation}
  \begin{array}{l}
  \ket{\psi} = \cos(\theta/2) \ket{0}_A \left( \cos(\theta_0/2) \ket{0}_B + \sin(\theta_0/2) e^{i\phi_0} \ket{1}_B \right) \\
  + \sin(\theta/2) e^{i\phi} \ket{1}_A \left( \cos(\theta_1/2) \ket{0}_B + \sin(\theta_1/2) e^{i\phi_1} \ket{1}_B  \right)
  \end{array}
\end{equation}

\section{Algorithm}\label{sec5}

In this section present the algorithm to go from the computational basis state representation to the binary tree of Bloch Sphere based on a recursive function. The following (trivially bijective mapping) is at the chore of the algorithm:

\begin{equation}
  \begin{array}{l}
     \forall \left( \phi_0,\phi_1 \in [0,2 \pi],r_0,r_1 \in \mathbb{R} \right)  \exists! \left( n \in \mathbb{R}, \theta \in [0,\pi], \phi_g,\phi_l \in [0,2 \pi] \right) s.t. \\ r_0 e^{i\phi_0} \ket{0} + r_1 e^{i\phi_1} \ket{1} = n e^{i\phi_g}( \cos{(\theta/2)} \ket{0} + \sin{(\theta/2)} e^{i\phi_l}) \ket{1}) 
  \end{array}
\end{equation}

In the above definition there exist a singularity in the case where $\theta$ is $0$ or $\pi$ in which case respectively $\phi_1$ or $\phi_0$ are not defined. Realising that it in fact correspond to a special case of product state where the parent is $\ket{0/1}$, it is clear that the correct mapping is to have the two subtrees equal, as in any product state. Therefore in that case we define the global phase corresponding to the remaining local phase $\phi_g = \phi_{0/1}$ and the local phase $\phi_l = 0$. The full recursive algorithm is presented in algorithm \ref{algo1}. It is efficient in the sense that it only queries each coefficient in the vector state once, which is the minimum queries required for a full representation.

We implemented the algorithm, together with visualisation tools in python. The BTBS figures included in this paper have been generated thanks to this code. It has been released as an open source library \emph{qutree} and is available on \href{https://github.com/alice4space/qutree}{GitHub} and \href{https://libraries.io/pypi/qutree}{PyPI}.

Discussion : The representation is very hierarchical in the sense that changing the order of qubits in which the Schmidt decomposition takes place changes completely the representation in a non-trivial way. 

\begin{algorithm}[H]
  \caption{Recursively compute the Bloch Sphere Binary tree}
  \begin{algorithmic}
    \State \textbf{Hyper-parameters} : The Schmidt qubit order 
    \State \textbf{Inputs} : $\ket{\psi} [2^N,M]$ $M$ samples of $N$ qubits vector states.
    \State \textbf{Outputs} : Data register (binary coordinate $b$, Bloch parameters $(\theta,\phi)$).
    \State \textbf{Helper function} Performs the function $f_h(r_0,\phi_0,r_1,\phi_1) = (n,\theta,\phi_l,\phi_g)$
    
    \\\hrulefill
    \State \textbf{Recursive function $F$}
    \\\hrulefill
    \State \textbf{Inputs} : 
    \begin{itemize}
  \item $\ket{\psi'} [2^{N'},M] $ tensor of state of subspace currently explored 
  \item $b$ Binary string of current coordinate, initialised empty (root)
  \item $D$ Data register, initialised empty
\end{itemize}

    \State \textbf{Outputs} : 
    \begin{itemize}
  \item $n$ the norm of the subspace explored
  \item $\phi$ the global phase to propagate up the tree
  \item $D$ Data register
\end{itemize}

    \State \textbf{Execute} :
\If{$N'=1$} \Comment{end of branch}
    \State $r_0 e^{i\phi_0}\gets \psi[0]$
    \State $r_1 e^{i\phi_1} \gets \psi[1]$

\Else \Comment{intermediate node}
    \State $\psi_0 \gets \psi'[:2^{N'-1}]$
    \State $\psi_1 \gets \psi'[2^{N'-1}:]$
    \State $r_0,\phi_0,D = F(\psi_0',b+"0",D) $ \Comment{recursive}
    \State $r_1,\phi_1,D = F(\psi_1',b+"1",D) $ \Comment{recursive}
\EndIf    
\State $(n,\theta,\phi_g,\phi_l) \gets f_h(r_0,\phi_0,r_1,\phi_1)$
\State $D \gets D + (b,\theta,\phi_l)$
\State \textbf{Return} $n,\phi_g,D$
 
  \end{algorithmic}
  \label{algo1}
\end{algorithm}

\pagebreak

\section{Applications}\label{sec3}

\subsection{Parameter Shift for Variational Quantum Algorithm}
\begin{figure}[H]
\centering
\includegraphics[width=0.9\linewidth]{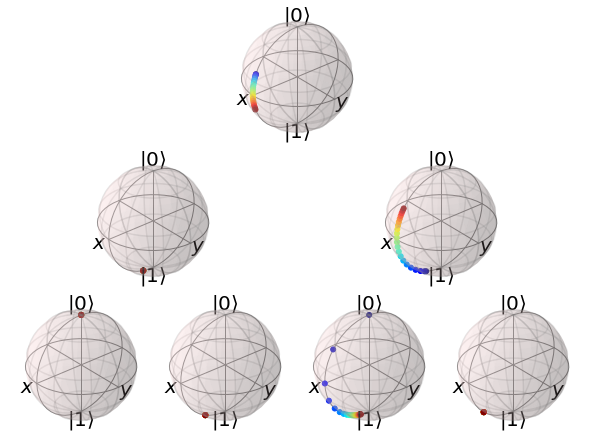}
\caption {amplitude basis encoding of $u$ where $u_5$ goes from 0 (blue) to 1 (red)}\label{figVQC}
\end{figure}

While crafting variational circuits the choice of architecture is challenging and the effect of certain parameters can remain elusive. The method described above allows to visualize the effect of shifting a parameter for a variational quantum circuit. In Figure \ref{figVQC} the states are an amplitude basis encoding $\ket{\psi} = \sum \ket{k} \frac{u_k}{\lVert u \rVert}$ of a random vector, with all coefficients are in $[0,1]$. We show the effect of varying the quantity $u_5$ from $t=0$ (blue) to $t=1$ (red).

\pagebreak

\subsection{Hamiltonian Time Evolution}

In this section we illustrate how this method can be used to visualise the Hamiltonian time evolution of quantum systems. We take a random hermitian matrix $H$.

\begin{figure}[H]
\centering
\includegraphics[width=0.9\linewidth]{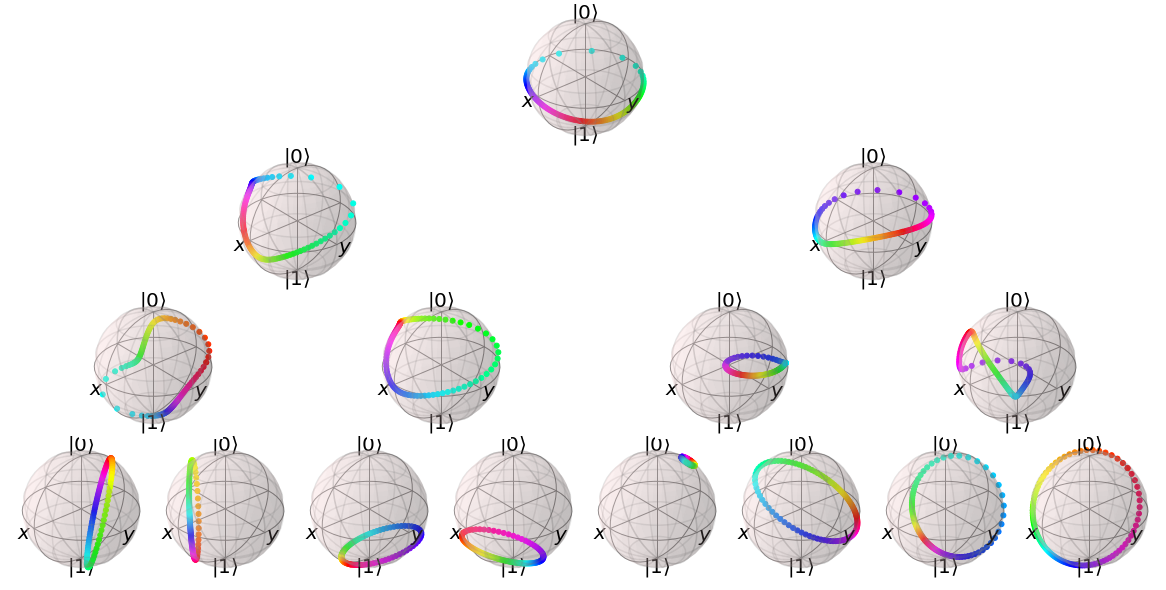}
\caption{Time Evolution with two excited eigenstates}\label{figTE1}
\end{figure}

First we simulate the time evolution starting from an initial condition exciting two eigenstates $\ket{\phi_1}$ $\ket{\phi_2}$ with corresponding eigenvalues $\lambda_1$ and $\lambda_2$. We know that the resulting time evolution will be periodic with period $\abs{\lambda_1 - \lambda_2}$. We show the resulting time evolution in \ref{figTE1}.

Secondly we simulate the same Hamiltonian from an initial condition exciting three eigenstates. The three corresponding eigenvalues don't have relative difference with rational ratios, therefore the time evolution will not be periodic. We show the resulting time evolution in \ref{figTE1}.

\begin{figure}[H]
\centering
\includegraphics[width=0.9\linewidth]{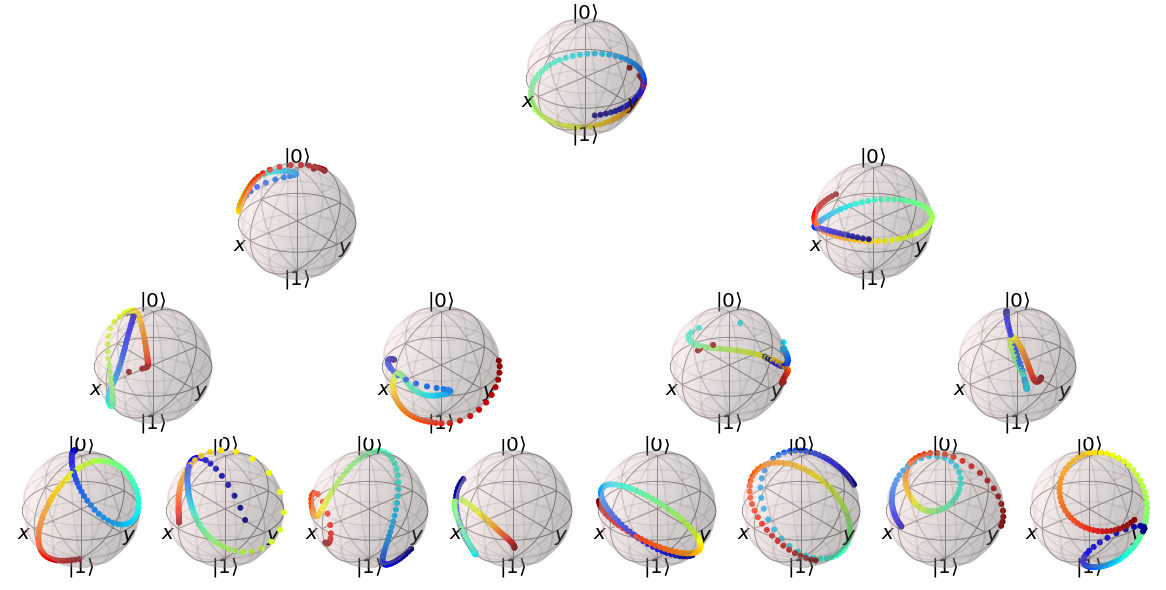}
\caption{Time Evolution with three excited eigenstates}\label{figTE2}
\end{figure}

\pagebreak

\subsection{Dataset visualisation}
\begin{figure}[H]
\centering
\includegraphics[width=0.9\linewidth]{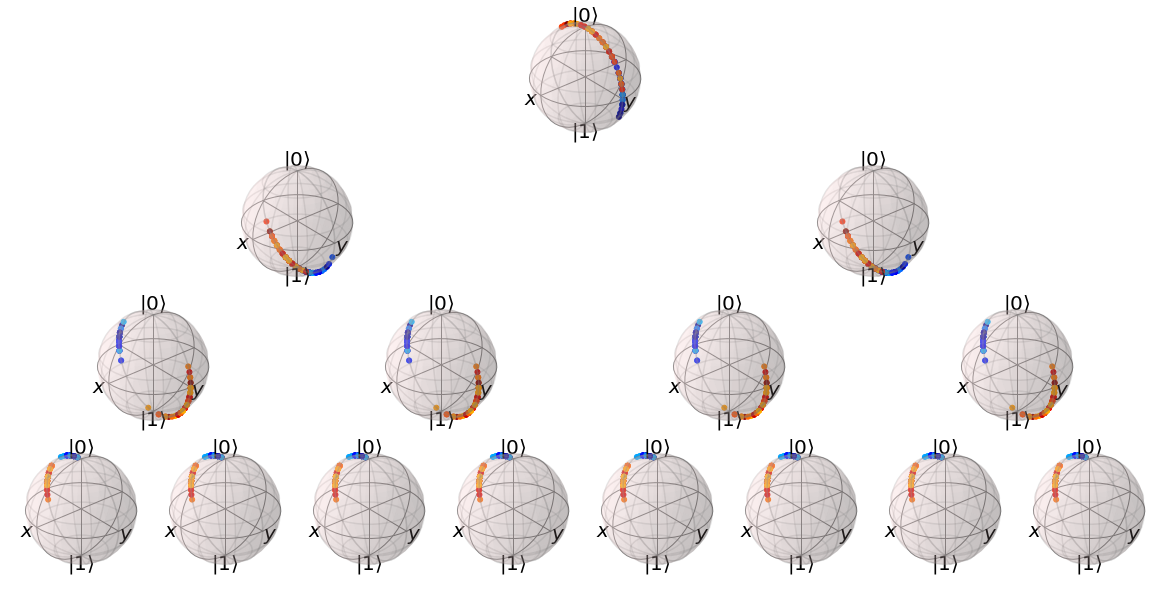}
\caption{Phase basis encoding}\label{figiris1}
\end{figure}

\begin{figure}[H]
\centering
\includegraphics[width=0.9\linewidth]{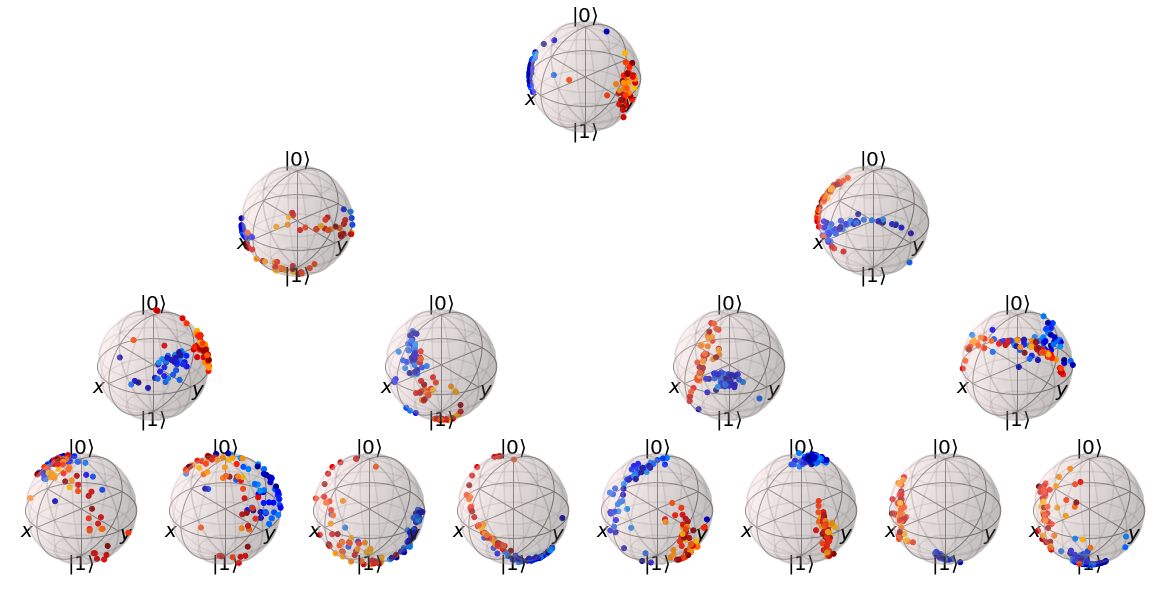}
\caption{Trained quantum kernel}\label{figiris2}
\end{figure}

Another application of the presented visualisation method is to look at quantum datasets. In this case we execute the demonstration of kernel learning applied to the iris dataset \cite{bib4} available in the documentation of the software library Pennylane developed by Xanadu. The dataset considered has four features and two linearly separable classes (blue and red). In the first part the classical data is uploaded with simple angle encoding; and then only the measurement is optimised to separate classes. On the visualisation \ref{figiris1}, it is possible to see that the encoding results in a product state, which we know it is the case per design, as spheres are copies of each others. Only a small portion of the Hilbert space is used. In the second part a quantum kernel is trained to separate the two classes based on the $Z$ expectation value of the first qubit. In the visualisation \ref{figiris2} it is possible to see that the resulting states are entangled and taking a much bigger portion of the Hilbert space, and that indeed the two classes are well separated.

\pagebreak

\end{document}